\documentclass[a4paper,12pt]{article}   %\linespread{1.3}
  \usepackage{makeidx}
  \usepackage{graphicx,graphics,color}
  \usepackage{latexsym,doc,fontenc,exscale}
 \usepackage{amsmath,amsfonts,amsthm,eucal,amssymb}

\begin{document}

 \title{\bf New Multi-order exact solutions for a class of nonlinear evolution equations}

\author{\textbf{Bijan Bagchi}$^{1}$\footnote{bbagchi123@rediffmail.com}, \textbf{Supratim
Das}$^{1}$\footnote{supratimiitkgp@gmail.com}
 and \textbf{Asish
 Ganguly}$^{2}$\footnote{gangulyasish@rediffmail.com}\\
$^{1}$\small{Department of Applied Mathematics, University of Calcutta},\\ \small 92 Acharya Prafulla Chandra Road, Kolkata-700009, India\\
 $^{2}$\small Department of Mathematics, Indian Institute of Technology, Kharagpur\\ \small Kharagpur-721302, India}
 \date{ }
\maketitle
  %\vspace{2mm}
 \noindent
 \hspace{-.8cm}Keywords: \textbf{\it  Travelling waves; Lam$\acute{e}$ equation; Lam$\acute{e}$ function; Jacobi elliptic function; Elliptic
 integral function.}
 \begin{abstract}
   We seek multi-order exact solutions of a generalized shallow water wave equation along with those corresponding
   to a class of nonlinear systems described by the KdV, modified KdV, Boussinesq, Klein-Gordon and modified
   Benjamin-Bona-Mahony equation. We employ a modified version of a generalized Lame equation and subject it
   to a perturbative treatment identifying the solutions order by order in terms of Jacobi elliptic functions. Our
   multi-order exact solutions are new and hold the key feature that they are expressible in terms of an auxiliary function $f$ in a
   generic way. For appropriate choices of $f$ we recover the previous results reported in the
   literature.\\
   %\vspace*{1.5cm}
 \noindent %
  \mbox{\hspace{-2.3cm}PACS number(s):  02.30.Jr, 02.30.Ik, 05.45.-a}
 \end{abstract}

 \section{Introduction}

 \hspace{0.5cm} Seeking tractable solutions of nonlinear evolution equations has been the focus of intense study for the past several
 decades (see, e.g. , \cite{cav}-\cite{gua}). While many strategies have been employed, such as the Hirota bilinear method \cite{hie1},
 homogeneous balance method \cite{wan},  trigonometric method \cite{yan}, the hyperbolic method \cite{her}, the Jacobi elliptic
 function method \cite{par} and iterative method \cite{daf1,daf2} to name a few,
perturbative techniques have also been used to extract multi-order
exact periodic solutions based
 on Lam$\acute{e}$ equation and Jacobi elliptic functions \cite{liu,fu,gua}. In this note we propose a generalized
 class of Lam$\acute{e}$ equation
 comprising an auxiliary function $f(\xi)$ to consider correlations between it and the perurbatively
 reduced nonlinear evolution equation for different orders of the perturbative parameter.

 A generalized Lam$\acute{e}$ equation is expressed as
 \begin{equation}\label{proposedODE}
 \frac{d^{2}y}{d\xi^{2}}+\left [\lambda-(n+1)(n+2)a_{1}a_{2} (f(\xi))^{2}\right ]y=0,
 \end{equation}
 where $\lambda$ stands for an eigenvalue, $n$ is a positive integer and we restrict the auxiliary function $f(\xi)$ to obey an elliptic equation
 \begin{equation}\label{elip}
 \left (\frac{df}{d\xi}\right )^{2}=\left ( 1+a_{1} f^{2}\right )\left ( a_{2}f^{2}+a_{3}\right ).
 \end{equation}
 In (\ref{proposedODE}) and (\ref{elip}), $a_{1},~a_{2},~a_{3}$ are real constants. It is well known that equation (\ref{elip}) admits of
 several categories of solutions for different values of $a_{1},~a_{2},~a_{3}$ all expressible in terms of modulus $k$ of the Jacobi elliptic functions.
 For later use we summarize a few relevant ones in Table~\ref{table1}.
 \begin{table}[ht]
 \caption{$f(\xi)$ is provided along with the values
 of the parameters $a_j\, , j=1,2,3$. The complementary modulus is denoted by $k'^{2}=1-k^{2}$ \cite{whi}.}
 \centering
\vspace*{2mm}
 \begin{tabular}{|c | c | c | c|}
 \hline
 $a_{1}$&   $a_{2}$ &  $a_{3}$ &  $f(\xi)$ \\[0.5ex]
 \hline
 $-1$ &  $-k^{2}$ & $1$ & $sn(\xi,k)$ \\
 $-1$ & $k^{2}$ & $k'^{2}$ & $cn(\xi,k)$ \\
 $-1$ & $1$ & $-k'^{2}$ & $dn(\xi,k)$ \\
 $-1$ & $-k'^{2}$ & $-k^{2}$ & $nc(\xi,k)~\mbox{i.e.}~\frac{1}{cn(\xi,k)}$ \\
 $-k'^{2}$ & $1$ & $-1$ & $nd(\xi,k)~\mbox{i.e.}~\frac{1}{dn(\xi,k)}$ \\
 $1$ & $k'^{2}$ & $1$ & $sc(\xi,k)~\mbox{i.e.}~\frac{sn(\xi,k)}{cn(\xi,k)}$ \\
 $-1$ & $-k^{2}$ & $1$ & $cd(\xi,k)~\mbox{i.e.}~\frac{cn(\xi,k)}{dn(\xi,k)}$ \\
 $-1$ & $-1$ & $k^{2}$ & $dc(\xi,k)~\mbox{i.e.}~\frac{dn(\xi,k)}{cn(\xi,k)}$ \\
  $k^{2}$ & $-k'^{2}$ & $1$ & $sd(\xi,k)~\mbox{i.e.}~\frac{sn(\xi,k)}{dn(\xi,k)}$ \\
 \hline
 \end{tabular}
 \label{table1}
 \end{table}

 It should be mentioned that while we would attend to the enlisted set of $f(\xi)$ in Table~\ref{table1} to widen the scope of
 enquiry, some particular cases of $f(\xi)$, namely $sn(\xi,k),~dn(\xi,k)$ and $cd(\xi,k)$, have been studied in the
 literature \cite{liu,fu,gua} to arrive at solutions of the relevant PDEs by applying the perturbation
 method on the latter. However, as is evident from Table~\ref{table1} there do remain other variants of $f(\xi)$ which open up
 different cases of new solutions not only for the already considered PDEs but other types as well by perturbatively
 reducing them to an ODE form so as to match with (\ref{proposedODE}).

  We have selected a physically meaningful model of a generalized shallow water wave equation \cite{hie,cla,che,bsa} which is
  amenable to a perturbative  treatment by looking for a travelling wave variable $u(x,t)=u(\xi)$ with $\xi=\gamma(x-ct)$ where $\gamma$ is the
  wave number and $c$ is the wave speed. We discuss the procedure of obtaining the solutions in the next section. In the following section we also
  take up applicability of our scheme to other  nonlinear evolution equations namely, the Korteweg de Vries (KdV) equation, modified KdV equation, Boussinesq
  equation, Klein-Gordon equation and  modified Benjamin-Bona-Mahony (mBBM) equation with a view to tracking down new hitherto unexplored multi-order
  solutions. The main point of our analysis is to demonstrate that all such solutions can be written down in a generic form in terms of the function $f$.

 To make equation (\ref{proposedODE}) accessible in terms of known Lam$\acute{e}$ functions, we recast it in terms of a new variable $\eta$ by
 applying the transformation $f(\xi)=\sqrt{\eta}$. Using (\ref{elip}) this gives
 \begin{equation}\label{lam}
 \frac{d^{2}y}{d\eta^{2}}+\frac{1}{2}\left [\frac{1}{\eta}+\frac{1}{\eta+\frac{1}{a_{1}}}
 +\frac{1}{\eta+\mu}\right ]\frac{dy}{d\eta}
 -\frac{\nu+(n+1)(n+2)\eta}{4\eta(\eta+\frac{1}{a_{1}})(\eta+\mu)}y=0,
 \end{equation}
 where the parameters $\mu$ and $\nu$ are defined by $\mu=\frac{a_{3}}{a_{2}}$ and $\nu=-\frac{\lambda}{a_{1}a_{2}}$.

 Equation (\ref{lam}) is readily solvable for the special cases of $n=1$ and $2$ for different choices of $\mu$ and $\nu$
 subject to certain relation between them. The solutions
 expressible by the corresponding Lam$\acute{e}$ functions are \cite{liu}
 \begin{eqnarray}\label{fun2}
 n&=&1~~~~~~~~~~L_{1}^{I}(\xi)=(1+a_{1}\eta)^{\frac{1}{2}}(a_{3}+a_{2}\eta)^{\frac{1}{2}}=(1+a_{1} f^{2})^{\frac{1}{2}}(a_{3}+a_{2}f^{2})^{\frac{1}{2}},~~\\
 &&~~~~~~\lambda=-(a_{2}+a_{1}a_{3})~~~[\mbox{provided}~\nu=(\mu+\frac{1}{a_{1}})].\nonumber
 \end{eqnarray}

 \begin{eqnarray}\label{fun3}
 n&=&1~~~~~~~~~~L_{1}^{II}(\xi)=\eta^{\frac{1}{2}}(1+a_{1}\eta)^{\frac{1}{2}}=f(1+a_{1} f^{2})^{\frac{1}{2}},~~~~~~~~~~~~~~~~~~~~~~~~~~~~~\\
 &&~~~~~~\lambda=-(a_{2}+4a_{1}a_{3})~~~[\mbox{provided}~\nu=(4\mu+\frac{1}{a_{1}})].\nonumber
 \end{eqnarray}

 \begin{eqnarray}\label{fun4}
 n&=&1~~~~~~~~~~L_{1}^{III}(\xi)=\eta^{\frac{1}{2}}(a_{3}+a_{2}\eta)^{\frac{1}{2}}=f(a_{3}+a_{2}f^{2})^{\frac{1}{2}},~~~~~~~~~~~~~~~~~~~~~~~~~\\
 &&~~~~~~\lambda=-(4a_{2}+a_{1}a_{3})~~~[\mbox{provided}~\nu=(\mu+\frac{4}{a_{1}})].\nonumber
 \end{eqnarray}

 \begin{eqnarray}\label{fun1}
 n&=&2~~~~~~~~~L_{2}(\xi)=\eta^{\frac{1}{2}}(1+a_{1}\eta)^{\frac{1}{2}}(a_{3}+a_{2}\eta)^{\frac{1}{2}}=
 f(1+a_{1} f^{2})^{\frac{1}{2}}(a_{3}+a_{2}f^{2})^{\frac{1}{2}},\\
 &&~~~~~~\lambda=-4(a_{2}+a_{1}a_{3})~~~[\mbox{provided}~\nu=4(\mu+\frac{1}{a_{1}})].\nonumber
 \end{eqnarray}

 Henceforth we will be guided by the solutions (\ref{fun2}) - (\ref{fun1}) along with an appropriate $f$ from Table 1 for the various
 multi-order cases that follow from a nonlinear PDE.
  %****************************************************************************
 \section{A generalized shallow water wave equation}\label{application}
 \hspace{0.5cm}We turn attention to a generalized class of shallow water wave (GSWW) equation given by \cite{hie,cla,che,bsa}
 \begin{equation}\label{GSWW}
 u_{xxxt}+\alpha u_{x}u_{xt}+\beta
 u_{t}u_{xx}-u_{xt}-u_{xx}=0,
 \end{equation}
 where $\alpha,~\beta\in\mathbb{R}-\{0\}.$ Under Boussinesq approximation the derivation of (\ref{GSWW}) results from the classical study of
 water waves. More interestingly, there also follows various classical and non-classical reductions from GSWW such as the KdV and BBM
 equations. Investigations of Painlev$\acute{e}$ tests reveal complete integrability for specific values of $\alpha$ and $\beta$
 namely $\alpha=\beta$ or $\alpha=2\beta$ \cite{cla}. Recently we extended \cite{bsa} the Jacobian elliptic function method to classify
 new exact travelling wave solutions expressible in terms of quasi-periodic elliptic integral function and doubly-periodic Jacobian elliptic functions.

 For the travelling wave solutions $u=u(\xi)$, equation (\ref{GSWW}) can be reduced
  in terms of the variable
 $v(\xi)\equiv\frac{du}{d\xi}$ to the ODE form
 \begin{equation}\label{rGSWW}
 \frac{d^{2}v}{d\xi^{2}}+Pv^{2}+Qv+c_{1}=0.
 \end{equation}
 where $P=\frac{\alpha+\beta}{2\gamma}$, $Q=\frac{1-c}{\gamma^{2}c}$ and $c_{1}$ is an integrating constant.

  To tackle (\ref{rGSWW}) perturbatively we set
 \begin{equation}\label{pert}
 v(\xi)=v_{0}(\xi)+\epsilon
 v_{1}(\xi)+\epsilon^{2}v_{2}(\xi)+\cdots
 \end{equation}
 where $\epsilon (>0)$ is a small parameter and $v_{0}(\xi),~v_{1}(\xi),~v_{2}(\xi),...$ represent various multi-order solutions like
 the zeroth-order, first-order, second-order solutions etc. of equation (\ref{rGSWW}). Accordingly we can write $u(\xi)$ as
 \begin{equation}\label{pert-u}
 u(\xi)=u_{0}(\xi)+\epsilon
 u_{1}(\xi)+\epsilon^{2}u_{2}(\xi)+\cdots
 \end{equation}
 where $u_{i}(\xi)=\int_{0}^{\xi}v_{i}(\tau)~d\tau.$

  Substituting the series (\ref{pert}) in equation (\ref{rGSWW}) we obtain for each power of $\epsilon$ the corresponding equations
 \begin{eqnarray}\label{reduODE}
 \epsilon^{0} &:& \frac{d^{2}v_{0}}{d\xi^{2}}+Pv_{0}^{2}+Qv_{0}+c_{1}=0\, , \\
 \epsilon^{1} &:& \frac{d^{2}v_{1}}{d\xi^{2}}+(2Pv_{0}+Q)v_{1}=0\, ,\label{f-ord-eq} \\
 \epsilon^{2} &:& \frac{d^{2}v_{2}}{d\xi^{2}}+(2Pv_{0}+Q)v_{2}=-Pv_{1}^{2}\, ,\label{s-ord-eq}
 \end{eqnarray}
 and so on.

 We now proceed to solve the above chain of equations by first expanding $v_{0}$ namely,
 \begin{equation}\label{z-ord}
 v_{0}=\sum_{i=0}^{l}A_{i}f^{i},
 \end{equation}
 where $A_{i}$'s are constants and then comparing highest order linear and nonlinear terms in (\ref{reduODE}). In this way we
 obtain $l=2$ by making use of (\ref{elip}). Thus (\ref{z-ord}) gets reduced to the quadratic form
 \begin{equation}\label{z-ord1}
 v_{0}=A_{0}+A_{1}f+A_{2}f^{2}.
 \end{equation}
 \subsection{Zeroth-order exact solution}
 \hspace{0.5cm} With $v_{0}$ given by (\ref{z-ord1}) we are led to the following system of coupled equations:
 \begin{eqnarray}\label{syst1}
 2a_{3}A_{2}+PA_{0}^{2}+QA_{0}+c_{1} &=& 0,\\
 A_{1}[(a_{2}+a_{1}a_{3})+2PA_{0}+Q] &=& 0\, ,\label{syst1-b}\\
 4(a_{2}+a_{1}a_{3})A_{2}+P(A_{1}^{2}+2A_{0}A_{2})+QA_{2} &=& 0\, ,\label{syst1-c} \\
 A_{1}[a_{1}a_{2}+PA_{2}] &=& 0 \, ,\label{syst1-d}\\
 A_{2}[6a_{1}a_{2}+PA_{2}] &=& 0.\, \label{syst1-e}
 \end{eqnarray}
 It is readily seen that we generate the following set of consistent solutions
 $A_{0}=-\frac{1}{2P}[Q+4(a_{2}+a_{1}a_{3})]$, $A_{1}=0$ and $A_{2}=-\frac{6a_{1}a_{2}}{P}$
 along with the constraint
 \begin{equation}\label{cons}
 [16(a_{2}+a_{1}a_{3})^{2}-48a_{1}a_{2}a_{3}]-Q^{2}+4Pc_{1}=0.
 \end{equation}
 From (\ref{cons}) we obtain wave speed as
 \begin{equation}
 c=[1\pm4\gamma^{2}\{(a_{2}+a_{1}a_{3})^{2}-3a_{1}a_{2}a_{3}+\frac{\alpha+\beta}{8\gamma}c_{1}\}^{\frac{1}{2}}]^{-1}
 \end{equation}
 where the two signs signal the two directions. Note that in order to have $c$ real we can always adjust the integration constant $c_{1}$ to keep $(a_{2}+a_{1}a_{3})^{2}+\frac{\alpha+\beta}{8\gamma}c_{1}>3a_{1}a_{2}a_{3}$. Further for the finiteness of $c$ we require $(a_{2}+a_{1}a_{3})^{2}+\frac{\alpha+\beta}{8\gamma}c_{1}\neq 3a_{1}a_{2}a_{3}+\frac{1}{16\gamma^{4}}$.

 From (\ref{reduODE}) and (\ref{z-ord1}) we get for the first integral
 \begin{equation}\label{v0}
 v_{0}=\frac{du_{0}}{d\xi}=-\frac{1}{2P}[Q+4(a_{2}+a_{1}a_{3})]-\frac{6a_{1}a_{2}}{P}f^{2}
 \end{equation}
 which in turn gives the zeroth-order solution
 \begin{equation}\label{z-ord-sol}
 u_{0}=-\frac{1}{2P}[Q+4(a_{2}+a_{1}a_{3})]\xi-\frac{6a_{1}a_{2}}{P}\int f^{2}~d\xi.
 \end{equation}

 As is evident, $u_{0}$ depends upon
 the choice of the auxiliary function $f$. In Table~\ref{table2} we furnish the various forms for $u_{0}$ corresponding to different
 elliptic functions of Table~\ref{table1}.
  %*****************************************Fig 1 (zero-exact.eps) & Fig 2(wave-speed.eps)******************************************************
\begin{figure}[ht]
\begin{center}
 \includegraphics[height=6cm,width=8cm]{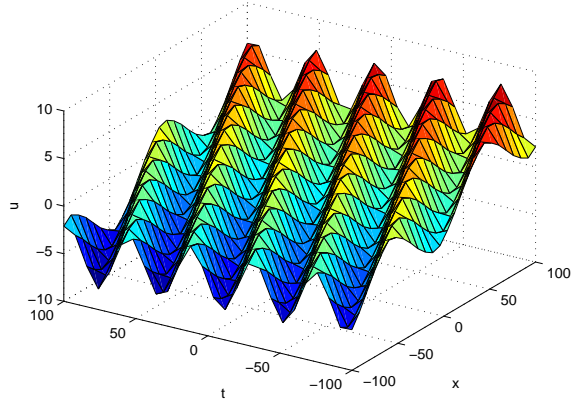}
 \end{center}
 \caption{Zeroth-order exact solution corresponding to $f(\xi)=sn(\xi,k)~\mbox{for}~k=0.5,~\alpha=1,~\beta=4,~\gamma=1$}\label{fig1}
\end{figure}
\begin{figure}[h]
\begin{center}
 \includegraphics[height=4cm,width=6cm]{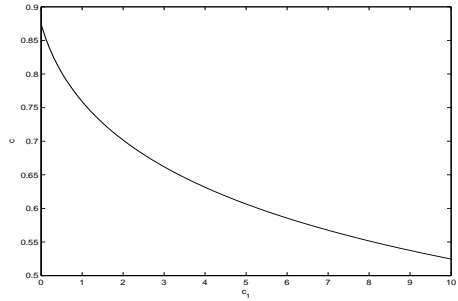}
 \end{center}
 \caption{Wave speed $c$ vs constant of integration $c_1$ for $f(\xi)=sn(\xi,k)~\mbox{for}~k=0.5,~\alpha=1,~\beta=4,~\gamma=1$ }\label{fig2}
\end{figure}
%**********************************************************************************************************
   \begin{table}[h]
 \caption{\small{Wave velocity $c=\frac{1}{1\pm4\gamma^{2}\sqrt{1-k^{2}k'^{2}+\frac{\alpha+\beta}{8\gamma}c_{1}}}$, $E(\phi,k)$ is the incomplete
 elliptic integral function of second kind
 where $sin\phi=sn(\xi,k)$ and $\Lambda=-\frac{1}{2P}[Q-4(1+k^{2})].$}}
 \centering
\vspace*{2mm}
 \small{\begin{tabular}{|c | c|}
 \hline
 $f(\xi)$ & Zeroth-order solution $u_{0}$ \\[0.5ex]
 \hline
 $sn(\xi,k)$ or $cn(\xi,k)$ or $dn(\xi,k)$ & $\Lambda\xi+\frac{6}{P}E(\phi,k)$ \\
 $nc(\xi,k)$ or $sc(\xi,k)$ or $dc(\xi,k)$ & $\Lambda\xi+\frac{6}{P}[E(\phi,k)-sn(\xi,k)dc(\xi,k)]$ \\
 $nd(\xi,k)$ or $sd(\xi,k)$ or $cd(\xi,k)$ & $\Lambda\xi+\frac{6}{P}[E(\phi,k)-k^{2}sn(\xi,k)cd(\xi,k)]$ \\
 \hline
 \end{tabular}}
 \label{table2}
 \end{table}
  A sample graph of the zeroth-order solution $u_{0}$ for $f=sn(\xi,k)$ is depicted in Figure~\ref{fig1} and a plot of wave-speed
  for variation of the constant $c_{1}$ is shown in Figure~\ref{fig2}.
 % \begin{figure}
%\begin{center}
 % \includegraphics[scale=0.4]{0.jpg}
 % \caption{Zeroth-order exact solution corresponding to $f(\xi)=sn(\xi,k)~\mbox{for}~k=0.5,~\alpha=1,~\beta=4,~\gamma=1$}\label{fig1}
 %\end{center}
%\end{figure}
%\begin{figure}
%\begin{center}
%  \includegraphics[scale=0.2]{c.jpg}
 % \caption{Wave speed $c$ vs constant of integration $c_1$ for $f(\xi)=sn(\xi,k)~\mbox{for}~k=0.5,~\alpha=1,~\beta=4,~\gamma=1$ }\label{fig2}
 %\end{center}
%\end{figure}
 %#############################################################################################################################
 \subsection{First-order exact solution}
 \hspace{0.5cm}Knowing $v_{0}$ from (\ref{v0}), equation (\ref{f-ord-eq}) reads
 \begin{equation}\label{r-f-ord-eq}
 \frac{d^{2}v_{1}}{d\xi^{2}}+[-4(a_{2}+a_{1}a_{3})-12a_{1}a_{2}f^{2}]v_{1}=0,
 \end{equation}
 which matches with (\ref{proposedODE}) for $\lambda=-4(a_{2}+a_{1}a_{3})$. This gives $n=2$ implying that the solution of (\ref{r-f-ord-eq}) can be
 written as
 \begin{equation}\label{v1}
 v_{1}=\frac{du_{1}}{d\xi}=c_{2}L_{2}(\xi)=c_{2}f(1+a_{1} f^{2})^{\frac{1}{2}}(1+\frac{a_{2}}{a_{3}}f^{2})^{\frac{1}{2}},
 \end{equation}
 where $c_{2}$ is an arbitrary constant.

\begin{figure}[ht]
\begin{center}
 \includegraphics[height=6cm,width=8cm]{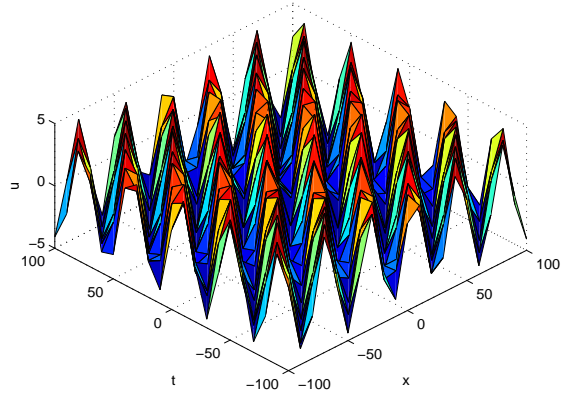}
 \end{center}
 \caption{First-order exact solution corresponding to $f(\xi)=sn(\xi,k)~\mbox{for}~k=0.5,~\alpha=1,~\beta=4,~\gamma=1,~c_{2}=1$}\label{fig3}
\end{figure}

 Integration of (\ref{v1}) with the use of (\ref{elip}) immediately provides
 \begin{equation}\label{f-ordsol}
 u_{1}=\pm\frac{c_{2}}{2}f^{2}
 \end{equation}
 which serves as a first order approximation to equation (\ref{GSWW}).
 Figure~\ref{fig3} gives a plot of $u_{1}$ for $f=sn(\xi,k)$.
 %*****************************************Fig 3 (first-exact.eps)******************************************************
%**************************************************************************************************************
% \begin{figure}[h]
%\begin{center}
%\includegraphics[scale=0.3]{1.jpg}
%\caption{First-order exact solution corresponding to $f(\xi)=sn(\xi,k)~\mbox{for}~k=0.5,~\alpha=1,~\beta=4,~\gamma=1,~c_{2}=1$}\label{fig3}
%\end{center}
%\end{figure}
%#######################################################################################################

 \subsection{Second-order exact solution}
 \hspace{0.5cm}Using (\ref{v0}) and (\ref{v1}) the second-order equation (\ref{s-ord-eq}) takes the form
  \begin{equation}\label{s-ord}
 \frac{d^{2}v_{2}}{d\xi^{2}}+[-4(a_{2}+a_{1}a_{3})-12a_{1}a_{2}f^{2}]v_{2}=-Pc_{2}^{2}f^{2}(1+a_{1}f^{2})(a_{3}+a_{2}f^{2}).
 \end{equation}
 We observe that the coefficients of (\ref{s-ord}) are polynomials in $f$. This prompts us to take a polynomial ansatz for $v_{2}$ namely
 \begin{equation}\label{v2-series}
 v_{2}=\sum_{i=0}^{l}B_{i}f^{i}.
 \end{equation}
 Substitution of the above into (\ref{s-ord}) we get $l=4$ on equating the highest power of $f$ from both sides. Thus (\ref{v2-series}) is
 reduced to the biquadratic form
 \begin{equation}\label{ans}
 v_{2}=B_{0}+B_{1}f+B_{2}f^{2}+B_{3}f^{3}+B_{4}f^{4}.
 \end{equation}

  Putting (\ref{ans}) into (\ref{s-ord}) and equating the coefficients of
 $f^{i}~(i=0~\mbox{to}~6)$ to zero, we have the following system of coupled equations,
 \begin{eqnarray}\label{syst2}
 a_{3}B_{2}-2(a_{2}+a_{1}a_{3})B_{0} &=& 0,\nonumber\\
 (a_{2}+a_{1}a_{3})B_{1}-2a_{3}B_{3}&=&0, \nonumber\\
 12a_{3}B_{4}-12a_{1}a_{2}B_{0}+Pc_{2}^{2}a_{3} &=& 0,\nonumber\\
 (a_{2}+a_{1}a_{3})B_{3}-2a_{1}a_{2}B_{1}&=&0,\\
 -6a_{1}a_{2}B_{2}+12(a_{2}+a_{1}a_{3})B_{4}+Pc_{2}^{2}(a_{2}+a_{1}a_{3}) &=& 0,\nonumber\\
 a_{1}a_{2}B_{3}&=&0,\nonumber\\
 8a_{1}a_{2}B_{4}+Pc_{2}^{2}a_{1}a_{2} &=& 0.\nonumber
 \end{eqnarray}
 Solving for $B_{0},~B_{1},~B_{2},~B_{3},~B_{4}$ we obtain
  \begin{equation}
 B_{0}=-\frac{a_{3}c_{2}^{2}P}{24a_{1}a_{2}},~B_{1}=0,~B_{2}=-\frac{(a_{2}+a_{1}a_{3})c_{2}^{2}P}{12a_{1}a_{2}},~B_{3}=0,
 ~B_{4}=-\frac{c_{2}^{2}P}{8}.
 \end{equation}
%*****************************************Fig 4 (second-exact.eps)******************************************************
\begin{figure}[ht]
\begin{center}
 \includegraphics[height=6cm,width=8cm]{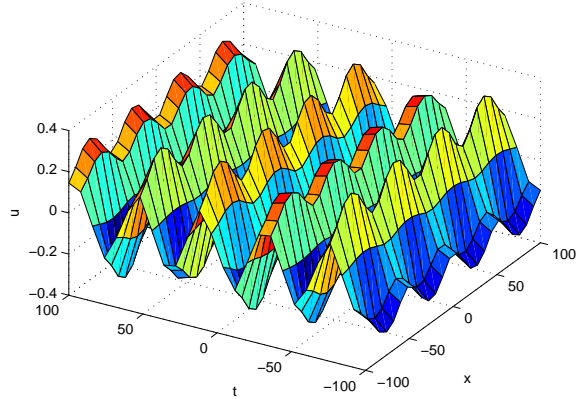}
 \end{center}
 \caption{Second-order exact solution corresponding to $f(\xi)=sn(\xi,k)~\mbox{for}~k=0.5,~\alpha=1,~\beta=4,~\gamma=1,~c_{2}=1$}\label{fig4}
\end{figure}
%**************************************************************************************************************

 Hence the solution of equation (\ref{s-ord}) can be written as
 \begin{equation}\label{v2}
 v_{2}=\frac{du_{2}}{d\xi}=-\frac{c_{2}^{2}P}{24a_{1}a_{2}}[a_{3}+2(a_{2}+a_{1}a_{3})f^{2}+3a_{1}a_{2}f^{4}]
 \end{equation}
 Using (\ref{elip}) we can integrate (\ref{v2}) to get
 \begin{equation}\label{GSWW-sos}
 u_{2}=\pm\frac{c_{2}^{2}P}{24a_{1}a_{2}}f(1+a_{1}f^{2})^{\frac{1}{2}}(a_{3}+a_{2}f^{2})^{\frac{1}{2}}.
 \end{equation}
 $u_{2}$ is plotted for $f=sn(\xi,k)$ in Figure~\ref{fig4}.
 %\begin{figure}
%\begin{center}
 % \includegraphics[scale=0.35]{2.jpg}
 % \caption{Second-order exact solution corresponding to $f(\xi)=sn(\xi,k)~\mbox{for}~k=0.5,~\alpha=1,~\beta=4,~\gamma=1,~c_{2}=1$}\label{fig4}
 % \end{center}
%\end{figure}

 %**************************************************************************
 \section{Application to other nonlinear \\evolution equations}\label{Summary}

 \hspace{.5cm} In the previous section, we considered a nonlinear second-order ODE containing a functional parameter $f$ and obtained
 specific solutions in terms of $f$. It was shown that the use of Jacobi elliptic function method in a perturbative way yielded new multi-order
 exact solutions that depended upon $f$ in a generic way. In this section we show that our approach can be profitably applied to other nonlinear
 equations resulting in more general types of solutions that have not been reported to the best of our knowledge. A complete list of our results
 are presented in Table~\ref{table3}.
  \small{ \begin{table}[ht]
 \caption{New general multi-order exact solutions for various non-linear evolution equations}
 \centering
\vspace*{2mm}
 \begin{tabular}{|c | c|}
 \hline
 Equation & Multi-order exact solutions $u_{n},~n=0,1,2$ \\[0.5ex]
 \hline
 $(i)$ KdV equation & $u_{0}=\frac{c}{\alpha}-\frac{4\beta\gamma^{2}}{\alpha}[(a_{2}+a_{1}a_{3})+3a_{1}a_{2}f^{2}],$\\
 $u_{t}+\alpha uu_{x}+\beta u_{xxx}=0$ & [where $c=\pm \beta\gamma^{2}\{16(a_{2}+a_{1}a_{3})^{2}-48a_{1}a_{2}a_{3}+2\frac{\alpha}{\beta\gamma^{2}}
                                                                                                          c_{1}\}^{\frac{1}{2}}$]\\
  & $u_{1}=c_{2}f(1+a_{1} f^{2})^{\frac{1}{2}}(a_{3}+a_{2}f^{2})^{\frac{1}{2}}$ \\
   & $u_{2}=-\frac{c_{2}^{2}\alpha}{48a_{1}a_{2}\beta\gamma^{2}}[a_{3}+2(a_{2}+a_{1}a_{3})f^{2}+3a_{1}a_{2} f^{4}]$ \\
 \hline
 $(ii)$ mKdV equation  & $u_{0}=\pm\sqrt{-\frac{6\beta\gamma^{2}a_{1}a_{2}}{\alpha}}~f,$\\
  $u_{t}+\alpha u^{2}u_{x}+\beta u_{xxx}=0$ & [where $c=\beta\gamma^{2}(a_{2}+a_{1}a_{3})$]\\
  & $u_{1}=c_{2}(1+a_{1}f^{2})^{\frac{1}{2}}(a_{3}+a_{2}f^{2})^{\frac{1}{2}}$\\
  & $u_{2}=\pm\frac{c_{2}^{2}}{2}\sqrt{-\frac{\alpha}{6\beta\gamma^{2}a_{1}a_{2}}}
 ~f[(a_{2}+a_{1}a_{3})+2a_{1}a_{2}f^{2}]$\\
 \hline
 $(iii)$ Boussinesq equation & $u_{0}=-\frac{1}{2\beta}[c_{0}^{2}-c^{2}+4\alpha\gamma^{2}(a_{2}+a_{1}a_{3})]-
 \frac{6\alpha\gamma^{2}}{\beta}a_{1}a_{2}f^{2},$\\
 $u_{tt}-c_{0}^{2}u_{xx}-\alpha u_{xxxx}-\beta (u^{2})_{xx}=0$ & [where $c=\{c_{0}^{2}\pm4\alpha\gamma^{2}
                                                               \sqrt{(a_{2}+a_{1}a_{3})^{2}-3a_{1}a_{2}a_{3}}\}^{\frac{1}{2}}$]\\
  & $u_{1}=c_{2}f(1+a_{1} f^{2})^{\frac{1}{2}}(a_{3}+a_{2}f^{2})^{\frac{1}{2}}$\\
   & $u_{2}=-\frac{c_{2}^{2}\beta}{24a_{1}a_{2}\alpha\gamma^{2}}[a_{3}+2(a_{2}+a_{1}a_{3})f^{2}+3a_{1}a_{2}f^{4}]$\\
 \hline
 $(iv)$ Klein-Gordon equation & $u_{0}=\pm\sqrt{-\frac{2(c^{2}-1)\gamma^{2}a_{1}a_{2}}{\beta}}~f,$\\
 $u_{tt}-u_{xx}+\alpha u+\beta u^{3}=0$ &  [where $c=\pm\{1-\frac{\alpha}{\gamma^{2}(a_{2}+a_{1}a_{3})}\}^{\frac{1}{2}}$]\\
  & $u_{1}=c_{2}(1+a_{1} f^{2})^{\frac{1}{2}}(a_{3}+a_{2}f^{2})^{\frac{1}{2}}$\\
  & $u_{2}=\pm\frac{c_{2}^{2}}{2}\sqrt{-\frac{\beta}{2(c^{2}-1)\gamma^{2}a_{1}a_{2}}}
 ~f[(a_{2}+a_{1}a_{3})+2a_{1}a_{2}f^{2}]$\\
 \hline
 $(v)$ mBBM equation & $u_{0}=\pm\sqrt{6\beta\gamma^{2}ca_{1}a_{2} }~f,$\\
  $u_{t}+c_{0}u_{x}+u^{2}u_{x}+\beta u_{xxt}=0$ & [where $c=\frac{c_{0}}{1+(a_{2}+a_{1}a_{3})\beta\gamma^{2}}$]\\
  & $u_{1}=c_{2}(1+a_{1}f^{2})^{\frac{1}{2}}(a_{3}+a_{2}f^{2})^{\frac{1}{2}}$\\
 & $u_{2}=\pm\frac{c_{2}^{2}}{2}\sqrt{\frac{1}{6\beta\gamma^{2}ca_{1}a_{2}}}~f[(a_{2}+a_{1}a_{3})
 +2a_{1}a_{2}f^{2}]$\\
 \hline
 \end{tabular}
 \label{table3}
 \end{table}}
 \newpage
  \textbf{$(i)$ KdV equation}\\

   The concerned equation is of the form
  \begin{equation}\label{kdv}
  u_{t}+\alpha uu_{x}+\beta u_{xxx}=0
  \end{equation}
  where $\alpha$ and $\beta$ are real parameters.
  Let us choose $f=cd(\xi,k)$ for which $a_{1}=-1,~a_{2}=-k^{2}~\mbox{and} ~a_{3}=1$. Substituting these values into the
  zeroth-order, first-order and second-order exact solutions listed in Table 3, we find
  \begin{eqnarray}\label{kdv-sol}
  u_{0}&=&\frac{c}{\alpha}+\frac{4\beta\gamma^{2}}{\alpha}(1+k^{2})-\frac{12\beta\gamma^{2}}{\alpha}k^{2}cd^{2}(\xi,k)\nonumber\\
  u_{1}&=&c_{2}k'^{2}cd(\xi,k)sd(\xi,k)nd(\xi,k)\\
  u_{2}&=&-\frac{c_{2}^{2}\alpha}{48\beta\gamma^{2}k^{2}}[1-2(1+k^{2})cd^{2}(\xi,k)+3k^{2}cd^{4}(\xi,k)].\nonumber
  \end{eqnarray}
  These solutions have been obtained in \cite{fu} for $\alpha = 1$ where the authors used the notations $m$ and $k$
 in places of $k$ and $\gamma$ respectively and their scale is $A=k'^{2}c_{2},~c_{1}=0$.
  The KdV equation has also been studied in \cite{liu} and their results can be extracted from ours for $f=sn(\xi,k)$
  corresponding to $\alpha=1$.\\\\\\
  \textbf{$(ii)$ mKdV equation}\\
  \begin{equation}
  u_{t}+\alpha u^{2}u_{x}+\beta u_{xxx}=0.
  \end{equation}
  Our general result listed in Table~\ref{table3} contains, as special cases, the following results reported in earlier works.
  \begin{itemize}
  \item $f=sn(\xi,k)$ (Ref \cite{liu})\\
  \begin{eqnarray}
  u_{0}&=&\pm\gamma k\sqrt{-\frac{6\beta}{\alpha}}~sn(\xi,k)\nonumber\\
  u_{1}&=&c_{2}~cn(\xi,k)dn(\xi,k)\\
  u_{2}&=&\pm\frac{(1+k^{2})c_{2}^{2}}{12\gamma k}\sqrt{-\frac{6\alpha}{\beta}}~sn(\xi,k)[\frac{2k^{2}}{1+k^{2}}~sn^{2}(\xi,k)-1]\nonumber
  \end{eqnarray}
  \item $f=cd(\xi,k)$ (Ref \cite{fu})\\
  \begin{eqnarray}
  u_{0}&=&\pm\gamma k\sqrt{-\frac{6\beta}{\alpha}}~cd(\xi,k)\nonumber\\
  u_{1}&=&(1-k^{2})c_{2}~sd(\xi,k)nd(\xi,k)\\
  u_{2}&=&\pm\frac{(1+k^{2})c_{2}^{2}}{12\gamma k}\sqrt{-\frac{6\alpha}{\beta}}~cd(\xi,k)[\frac{2k^{2}}{1+k^{2}}~cd^{2}(\xi,k)-1]\nonumber
  \end{eqnarray}
  \end{itemize}
  \vspace{0.4cm}
  \textbf{$(iii)$ Boussinesq equation}\\
  \begin{equation}
  u_{tt}-c_{0}^{2}u_{xx}-\alpha u_{xxxx}-\beta (u^{2})_{xx}=0.
  \end{equation}
  From the general solution given in terms of $f$ (see Table~\ref{table3}) one can deduce the ones obtained
  in  \cite{liu} as special case for $f = sn(\xi,k)$.\\\\\\
  \textbf{$(iv)$ Klein-Gordon equation}\\
  \begin{equation}
  u_{tt}-u_{xx}+\alpha u+\beta u^{3}=0.
  \end{equation}
  The above equation was considered in \cite{gua}. Their results correspond to the case $f=dn(\xi,k)$ and we do indeed recover
  their solutions from Table 3 as given below:
  \begin{eqnarray}
  u_{0}&=&\pm\sqrt{\frac{2\alpha}{\beta\gamma^{2}(k^{2}-2)}}~dn(\xi,k)\nonumber\\
  u_{1}&=&k^{2}c_{2}~sn(\xi,k)cn(\xi,k)\\
  u_{2}&=&\mp\frac{c_{2}^{2}}{2}\sqrt{\frac{\beta}{2\gamma^{2}(c^{2}-1)}}~dn(\xi,k)[2-k^{2}-2dn^{2}(\xi,k)]\nonumber
  \end{eqnarray}
  \vspace{0.4cm}
  \textbf{$(v)$ mBBM equation}\\
  \begin{equation}\label{mbbm}
  u_{t}+c_{0}u_{x}+u^{2}u_{x}+\beta u_{xxt}=0.
  \end{equation}
  For $f=sn(\xi,k)$ the multi-order solutions to (\ref{mbbm}) read from Table 3
  \begin{eqnarray}
  u_{0}&=&\pm\gamma\sqrt{6c\beta}~sn(\xi,k)\nonumber\\
  u_{1}&=&c_{2}~cn(\xi,k)dn(\xi,k)\\
  u_{2}&=&\mp\frac{c_{2}^{2}(1+k^{2})}{12\gamma k}\sqrt{\frac{6}{c\beta}}~sn(\xi,k)[12-\frac{2k^{2}}{1+k^{2}}sn^{2}(\xi,k)].\nonumber
  \end{eqnarray}
  These results were obtained in Ref~\cite{liu}.

  %########################################################################################################################################
  \section{Concluding remarks}
  \hspace{0.5cm} We have reported exact multi-order periodic solutions for a generalized shallow water wave
  equation based on various types of Jacobi elliptic functions in a perturbative framework. The presence of an auxiliary
  function, which makes reference to any particular Jacobi function explicit, can be well adjusted to connect with other
  classes of PDE such as the KdV, modified KdV, Boussinesq, Klein-Gordon and modified Benjamin-Bona-Mahony equation. The
  general character of our results encompasses those special cases which have earlier been studied in the literature.

%\newpage

 \end{document}